\documentclass[prb,aps,preprint]{revtex4}
\usepackage{amsmath}
\usepackage{graphicx}
\begin{document}

\title{Stress modulated optical spin-injection in bulk semiconductors}
\author{Jos\'e Luis Cabellos}\email[email:]{sollebac@gmail.com}
\author{Cuauht\'emoc Salazar}
\author{Bernardo S. Mendoza}
\affiliation{Division of Photonics,
Centro de Investigaciones en \'Optica,
Loma del Bosque 115,
Le\'on, Guanajuato, M\'exico}
\date{\today}

\begin{abstract}
A full band-structure \textit{ab initio} calculation of the degree of
spin polarization (DSP) in stressed bulk Si and bulk GaAs is reported.
For Si, we found that compressive stress causes the DSP signal peak to
decrease slightly in magnitude and to shift to higher energies.  For
expansive stress, the DSP signal shows a notable enhancement,
changing from -31.5\% for the unstressed case to +50\% with only 1.5\%
of volumetric change.
For GaAs, the only change induced due to either expansive or
compressive stress is an energy shift of the DSP spectrum. This
behaviour may serve to tune the DSP in semiconductors to a suitable laser
energy.
\end{abstract}

\maketitle

The study of spin injection into a non magnetic semiconductor is an
important field of research in condensed matter physics, known as
spintronics, which has the potential of many
applications.\cite{zuticRMP04} The optical excitation of
semiconductors with circularly polarized light creates spin-polarized
electrons in the conduction bands.\cite{dyakonovE84} The idea of
using light for spin injection and detection dates back to 
1968.\cite{lampelPRL68} Later it was shown that conversion of
angular momentum of light into electron spin and vice versa is very
efficient in III-IV semiconductors.\cite{dyakonovE84} Known as
``optical orientation'', this effect serves as an important tool in
the field of spintronics, where it is used to spin-polarize electrons.
The injection of spin and the degree of spin polarization in bulk
GaAs, Si and CdSe semiconductors has been reported
recently,\cite{nastosPRB07} where a detailed comparison between a
30-band $\mathbf{k}\cdot\mathbf{p}$ model and a full band structure LDA (local
density approximation) + scissors correction calculation was given.
Some of the results obtained could be explained simply by using
well-known features of the band structure and selection rules around
the $\Gamma$-points of GaAs and Si.  However, for photon energies well
above the band gap the selection rules are more complicated and full
band structure calculations are required to explore the degree of spin
polarization.  For many semiconductors, like CdSe, no $\mathbf{k}\cdot\mathbf{p}$ models are
available, and the results of Nastos et al.\cite{nastosPRB07} indicate
that the degree of spin polarization can be reliably calculated with
LDA + scissors corrected band structures.  This suggests a program of
study of optical orientation based on LDA + scissors
calculations. Recently Cabellos et al.\cite{cabellosPRB09c} have
extended such theoretical study to several Si(111) surfaces, founding 
that these surfaces exhibit a degree of spin polarization (DSP) larger than the bulk Si DSP.

The purpose of this work is to calculate the DSP in stressed bulk Si
and stressed bulk GaAs.  We characterized applied stress by isometric
volumetric strains, where the ratio of the volume at the stressed state
to the volume at the unstressed state is employed as the independent
input-variable.  We compute the DSP for a set of volumetric
strains. To avoid structural changes, which are reported to arise
at about 10\% of volumetric change,\cite{turneaureComm} we restricted our
computations between the range of  $1.5\%$ of expansive  strain and $-1.5\%$ of
compressive strain.

The theory of DSP is laid out by Nastos et al.,\cite{nastosPRB07} where
we refer the reader for the details. Here, we only
reproduce the most important expressions in order to calculate the
DSP. The DSP along direction ``a'' is formally defined as
\begin{eqnarray}\label{eq:dsp}
 {\cal D}^{\mathrm{a}} =\frac{\dot S^{\mathrm{a}}}{(\hbar/2)\dot n}
, 
\end{eqnarray}
where the rate of spin injection is given by
$\dot S^{\mathrm{a}}= \zeta^{\mathrm{abc}}(\omega)
E^{\mathrm{b}}(-\omega) E^{\mathrm{c}}(\omega)
$ 
and the rate of carrier injection by
$\dot n=\xi^{\mathrm{ab}}(\omega)
E^{\mathrm{b}}(-\omega) E^{\mathrm{c}}(\omega)
$. 
Also,
\begin{eqnarray}\label{zetaabci}
\zeta^{\mathrm{abc}}(\omega)
&=&
\frac{i\pi e^2}{\hbar^2}
\int\frac{d^3k}{8\pi^3}
\sum_{vcc'}\,'\,
\mathrm{Im}\Big[S^{\mathrm{a}}_{c'c}(\mathbf{k}) r^{\mathrm{b}}_{vc'}(\mathbf{k}) r^{\mathrm{c}}_{cv}(\mathbf{k})
\nonumber\\
&+& 
S^{\mathrm{a}}_{cc'}(\mathbf{k}) r^{\mathrm{b}}_{vc}(\mathbf{k}) r^{\mathrm{c}}_{c'v}(\mathbf{k})\Big]
 \delta(\omega_{cv}(\mathbf{k})-\omega)
,
\end{eqnarray}  
is the (purely imaginary)  pseudo-tensor that allows us to 
calculate the spin injection rate, and
\begin{eqnarray}\label{imxi}
\xi^{\mathrm{ab}}(\omega) 
&=&
\frac{2\pi e^2}{\hbar^2}
\int\frac{d^3k}{8\pi^3}
\sum_{vc}
\mbox{Re}[
r^{\mathrm{a}}_{vc}(\mathbf{k})
r^{\mathrm{b}}_{cv}(\mathbf{k})
]
\nonumber\\
&\times&
\delta(\omega_{cv}(\mathbf{k})-\omega)
,
\end{eqnarray}
is the tensor that allows us to calculate the carrier injection.  The
roman Cartesian superscripts are summed over if repeated. The above
results take into account the coherent processes that take place in the
spin-split conduction bands due to the finite width of the laser
pulse. For this reason  the prime in the sum of Eq.~\eqref{zetaabci} is
restricted to conduction bands $c$ and $c'$ that are closer than 30
meV.\cite{nastosPRB07} The matrix elements of the position operator
$r^{\mathrm{a}}_{nm}(\mathbf{k})$, the spin operator
$S^{\mathrm{a}}_{nm}(\mathbf{k})$,
 and the
energy difference between valence ($v$) and conduction ($c$) states,
$\omega_{cv}(\mathbf{k})$, are evaluated for $\mathbf{k}$-points on a specially determined
tetrahedral grid.  This grid is used in the integrals of
Eqs.~\eqref{zetaabci} and~\eqref{imxi} that are calculated through a
linear analytic tetrahedral integration method.\cite{nastosPRB07} We
assume, as is commonly done,\cite{dyakonovE84} that the hole spins
relax very quickly and we neglect them, focusing only on the electron
spins; measurements have led to estimates of 110~fs for the heavy-hole
spin life time in GaAs.\cite{hiltonPRL02}

The calculations were performed in the framework of the density
functional theory (DFT) with the local density approximation (LDA) +
scissors correction, using the ABINIT plane-wave code.\cite{gonzeCMS02}
To include the spin-orbit interaction, we use the separable
Hartwigsen-Goedecker-Hutter pseudopotentials\cite{hartwigsenPRB98}
within the LDA as parametrized by Goedecker {\it et
  al.}\cite{goedeckerPRB96} We exclude the semi-core states (though
they can be included with more computational effort), the
contributions to the velocity matrix elements from the nonlocal part
of the pseudopotential and from the spin-orbit interaction.  However,
we know that the contributions of the last two are small for
Si.\cite{mendozaPRB06,readPRB91,kageshimamPRB97} The scissors
correction amounts to a rigid shift of ${\cal D}^a$ along the energy axis
by 0.87 eV (1.16 eV) for Si (GaAs) that is the value required to
increase the LDA gap at the $\Gamma$ point to its experimental
value.\cite{cabellosPRB09a,nastosPRB05} Since the core electrons are
neglected, we have 8 electrons with spin up and spin down wave
functions and thus 8 valence bands.  Consequently we found converged
results with just 8 conduction bands, along with a cutoff of
30~Hartree and 18424 $\mathbf{k}$-points.

For Si and GaAs their corresponding crystal classes
 have the following non-zero components: 
$\zeta^{\mathrm{ zxy}}=\zeta^{\mathrm{
    yzx}}=\zeta^{\mathrm{ xyz}} =-\zeta^{\mathrm{
    zyx}}=-\zeta^{\mathrm{ yxz}}=-\zeta^{\mathrm{ xzy}}$, and
$\xi^{\mathrm{ xx}}=\xi^{\mathrm{ yy}}=\xi^{\mathrm{ zz}}\equiv\xi$.
Using a circularly
left-polarized electric field propagating along the $-z$ direction, 
i.e.
$\mathbf{E}=E_0 \big( \hat{\mathbf{x}} - i\hat{\mathbf{y}} \big)/\sqrt{2}$
with $E_0$ its
intensity, we get 
from Eq.~\eqref{eq:dsp} 
 the DSP along the direction
of propagation of the electric field as
$ {\cal D}^\mathrm{ z} =\zeta^{\mathrm{ zxy}}/ (\hbar\xi/2) $.
We characterized the applied
stress by isometric volumetric strains. Then, the ratio of the volume
at the stressed state, $\Omega_s=a_s^3$, to the volume at the unstressed
state, $\Omega_0=a_0^3$, is given by $\sigma=a_s/a_0$, where $a_0=5.39$~\AA 
(5.53~\AA), is
the unstressed lattice parameter of the cubic unit cell of Si (GaAs), and
$a_s$ is the stressed value.  We use $a_s=\sigma a_0$ as the independent
variable to calculate ${\cal D}^{\mathrm{ z}}$ vs. $\sigma$.

\begin{figure}[t]
\includegraphics[scale=0.9]{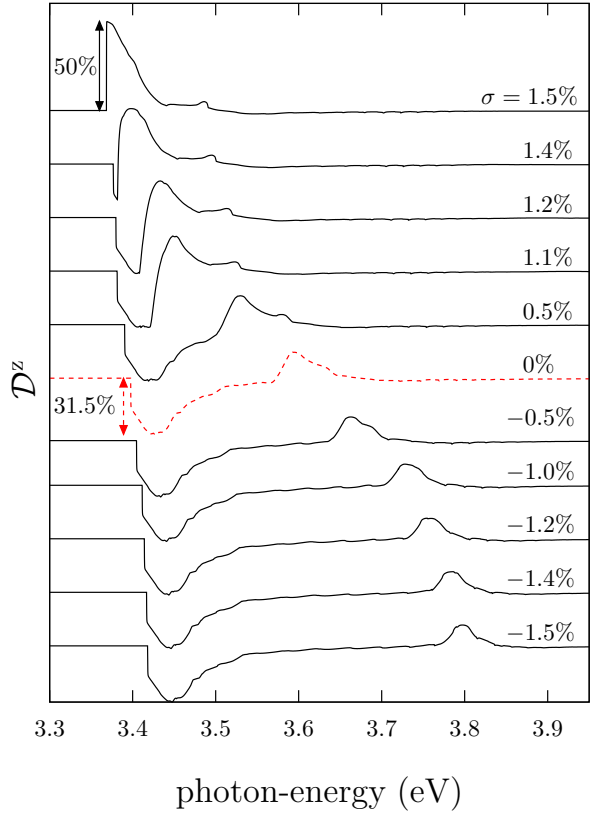}
\caption{(color online) 
Stress
modulation of the DSP, ${\cal D}^{\mathrm{ z}}$, vs. photon energy for bulk silicon. 
Several spectra for different values of $\sigma$ (expressed as 
percentage) are shown, where $\sigma>0$ ($\sigma<0$) is for
  expansive  (compressive) stresses.
The unstressed ${\cal D}^{\mathrm{ z}}$ ($\sigma=0\%$) is shown by a dotted line with a maximum
value of $|{\cal D}^{\mathrm{ z}}|=31.5\%$. For $\sigma=1.5\%$,
${\cal D}^{\mathrm{ z}}|_{\mathrm{max}}=50\%$. Each spectrum has been offset in the
vertical axis for displaying purposes.
}
\label{fig_dsp_si}
\end{figure}

For Si we show in
Fig.~\ref{fig_dsp_si} the calculated
${\cal D}^{\mathrm{ z}}$ vs. the photon energy for several values of
$\sigma$, including both expansive and compressive strains,  along with the result of
the unstressed ($\sigma=0$) result. We vary $\sigma$ from $-1.5\%$ to 1.5\%.
The unstressed spectrum shows two main features, one at 3.425 eV,
just a few meV above the band gap with a $-31.5\%$ deep, and the other
at 3.593~eV with a $15\%$ peak. As we compress the unit cell
($\sigma<0$) we see that the negative deep remains almost unchanged in
magnitude and energy position, however the positive peak moves towards
higher energies, keeping almost the same shape and showing a modest 
reduction to 11\% at $\sigma=-1.5\%$.  This situation changes radically
when we expand the unit cell.  Indeed, as $\sigma$ increases the negative
deep gets narrower, slightly moves to lower energies and then disappears
at $\sigma=1.403\%$.  The positive peak in turn moves to lower energies,
increases its height and its shape changes until it gives a
${\cal D}^{\mathrm{ z}}$ that rises sharply at the band edge with a
maximum intensity of 50\%. The spectrum at $\sigma=1.5\%$ only shows
this positive peak that has the largest $|{\cal D}^{\mathrm{ z}}|$ magnitude
of all the spectra.
Thus, under
expansive stress bulk Si exhibits a quite interesting response: the
negative deep and positive peak shown in ${\cal D}^{\mathrm{ z}}$ for the
unstressed unit cell coalesce into a single positive peak at the band
edge with $50\%$ of the spins polarized along the direction of propagation of
the optical beam.  We have checked that for $\sigma>1.5\%$, the
${\cal D}^{\mathrm{ z}}$ only shifts the spectrum to lower energies,
retaining the magnitude of the DSP signal peak at $50\%$. Nevertheless
such large expansions may be experimentally more difficult to
achieve.\cite{notamty}

\begin{figure}[t]
\includegraphics[scale=0.9]{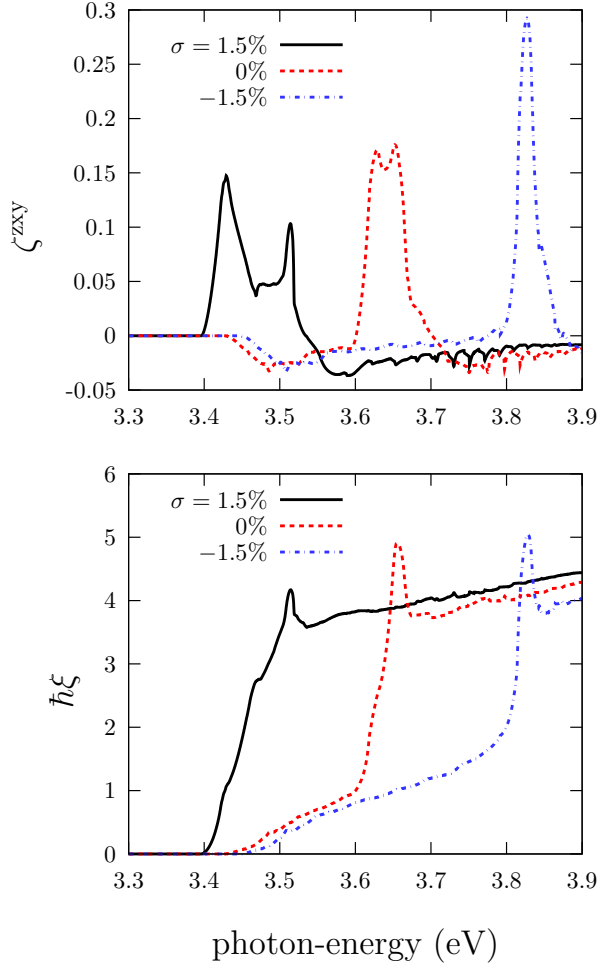}
\caption{(color online) 
The calculated $\zeta^{\mathrm{ zxy}}(\omega)$ (top panel) 
and
$\hbar\xi(\omega)$
(bottom panel) 
vs. photon energy for three different values of strain ($\sigma$).
}
\label{fig_zx_si}
\end{figure}

\begin{figure}[t]
\includegraphics[scale=0.9]{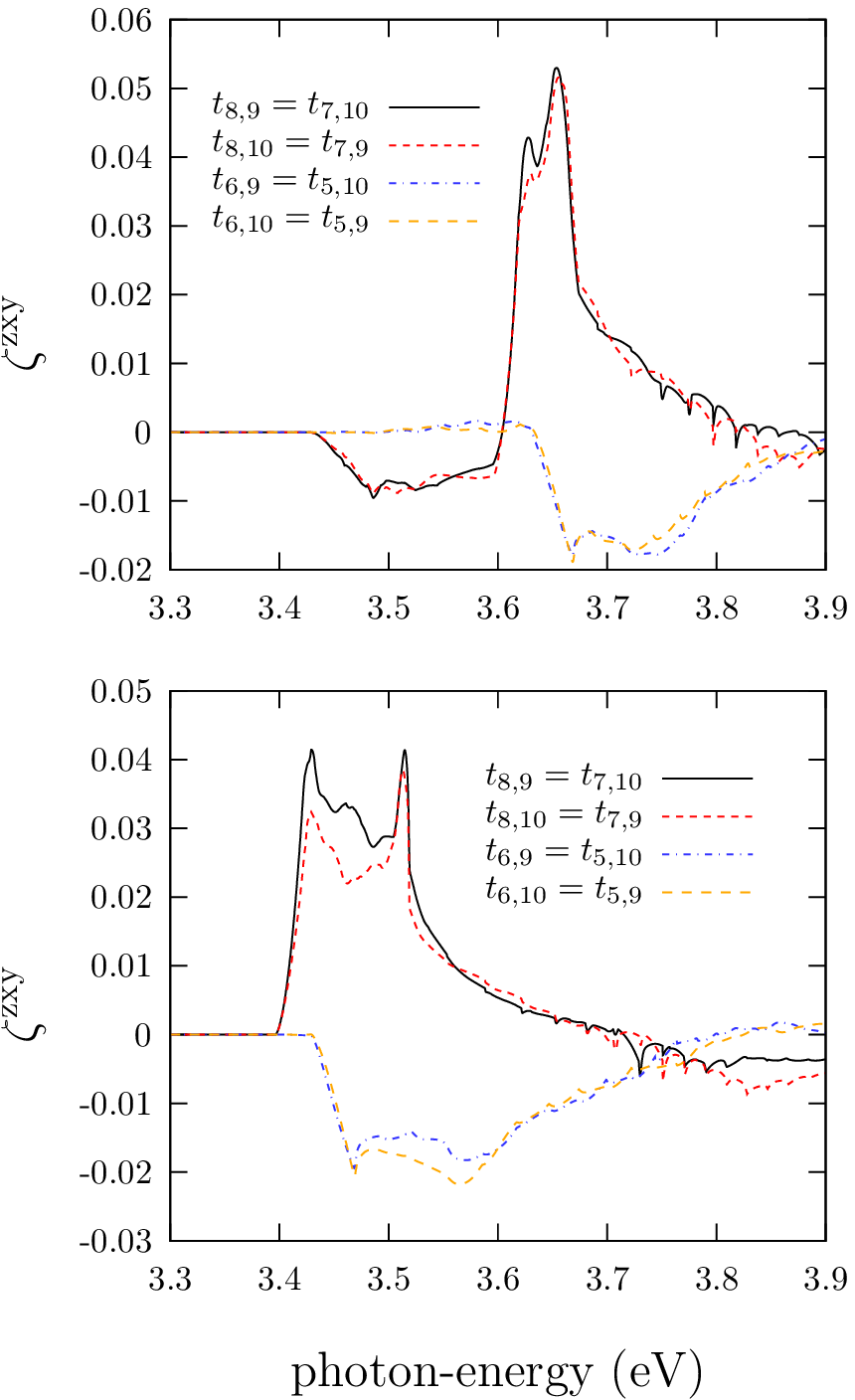}
\caption{(color online)
Breakdown of the rate of spin injection, $\zeta^{\mathrm{zxy}}(\omega)$,
 into different band
contributions
for $\sigma=0$ (top panel) and $\sigma=1.5\%$ (bottom panel).
We use $t_{m,n}=v_m\to c_n$ to denote a transition from 
the $v_m$-valence band to the $c_n$-conduction band (see text for details).
}
\label{fig_vc-zeta_si}
\end{figure}

In Fig.~\ref{fig_zx_si} we show the calculated $\zeta^{\mathrm{
    zxy}}(\omega)$ and $\hbar\xi(\omega)$ 
for $\sigma=0,\pm1.5\%$.  
We remark that in Gaussian units both tensors are dimensionless
quantities. 
We see that the onset
at the band-edge is red-shifted in energy as $\sigma$ goes from $-1.5\%$ to
1.5\%. For both $\sigma=0$ and $-1.5\%$ $\zeta^{\mathrm{ zxy}}(\omega)$ is negative
around the onset, whereas it is positive for $\sigma=1.5\%$ and rises very
sharply.  For $\hbar\xi(\omega)$ we see that the rise of the signal at the
onset changes also with $\sigma$, being rather sharp for $\sigma=1.5\%$ as
it is for $\zeta^{\mathrm{ zxy}}(\omega)$. From these results, one can
understand the line shape of ${\cal D}^{\mathrm{ z}}$ shown in
Fig.~\ref{fig_dsp_si}. Indeed, the minimum (maximum) present in
${\cal D}^{\mathrm{ z}}$ for $\sigma=0,-1.5\%$, comes from the minimum
(maximum) in $\zeta^{\mathrm{ zxy}}(\omega)$, whereas the only one maximum of
${\cal D}^{\mathrm{ z}}$ for $\sigma=1.5\%$ near the band-edge comes from
the maximum at $\zeta^{\mathrm{ zxy}}(\omega)$, but then the next local
maximum in $\zeta^{\mathrm{ zxy}}(\omega)$ is barely seen in 
${\cal D}^{\mathrm{z}}$
 since, as shown in the Fig.~\ref{fig_zx_si}, the
corresponding $\hbar\xi(\omega)$ is rather large as compared with 
$\zeta^{\mathrm{zxy}}(\omega)$.
 In other words, the DSP depends strongly on the fine
interplay between the ability for polarizing the spin of the
electrons, and the number of electrons (carriers) that one can inject.

The rate of spin injection from the different transitions in our
calculation can be seen more clearly in Fig.~\ref{fig_vc-zeta_si},
where we show $\zeta^{\mathrm{zxy}}(\omega)$ for selected transitions from the
valence bands ($v$) to the conduction bands ($c$), for $\sigma=0$ and
$\sigma=1.5\%$.  We only show the transitions that have the most
influential effect
on the net spin-injection rate.  We label the bands in ascending values of 
energy, thus $v_8$ is the top valence band and $c_9$ is the bottom
conduction band.  We use $t_{m,n}=v_m\to c_n$ to denote a transition
from $v_m$ ($m\le 8$) to $c_n$ ($n\geq 9$).  Indeed, we see how the
onset of the response is mainly given by the contributions of
$t_{8,9}$, $t_{8,10}$, $t_{7,9}$, $t_{7,10}$, $t_{6,9}$, $t_{6,10}$,
$t_{5,9}$, and $t_{5,10}$, where the following relationships,
traceable to the symmetry of the states,\cite{nastosPRB07} are found:
$t_{8,9}=t_{7,10}$, $t_{8,10}=t_{7,9}$, $t_{6,9}=t_{5,10}$, and
$t_{6,10}=t_{5,9}$.  These are the transitions from the last four
valence bands to the bottom of the first two conduction bands.  We see
that the onset of the signal at the corresponding band edge for each
value of $\sigma$ is due to the $t_{8,9}=t_{7,10}$ and the
$t_{8,10}=t_{7,9}$ transitions.  These transitions for $\sigma=0$ have a
$\zeta^{\mathrm{zxy}}(\omega)$ that is first negative from 3.40~eV till
3.58~eV, and then becomes positive and goes to almost zero above
3.86~eV.  However, for the same transitions at $\sigma=1.5\%$, the corresponding
$\zeta^{\mathrm{zxy}}(\omega)$ is always positive
and  goes to almost zero above 3.9~eV.  Also, the spectra shows
that the $t_{6,9}=t_{5,10}$ and the $t_{6,10}=t_{5,9}$ transitions
kick in above the band gap and that for both values of $\sigma$ the
corresponding $\zeta^{\mathrm{zxy}}(\omega)$ is always negative.  But now we
note that, for $\sigma=0$, these transitions kick in 200~meV above the
band gap, whereas for $\sigma=1.5\%$ the signal kick in just 32~meV above
the band gap.  This large difference in turn gives the
${\cal D}^{\mathrm{z}}$ observed in Fig.~\ref{fig_dsp_si}, i.e. for
$\sigma=0$ we have a broad minimum at 25~meV above the band edge, followed
by a broad maximum at 195~meV above the band edge, whereas for
$\sigma=1.5\%$ we have a sudden build up of ${\cal D}^{\mathrm{z}}$ at the
band edge followed by a rapid decrease of the signal to zero.

\begin{figure}
\includegraphics[scale=0.9]{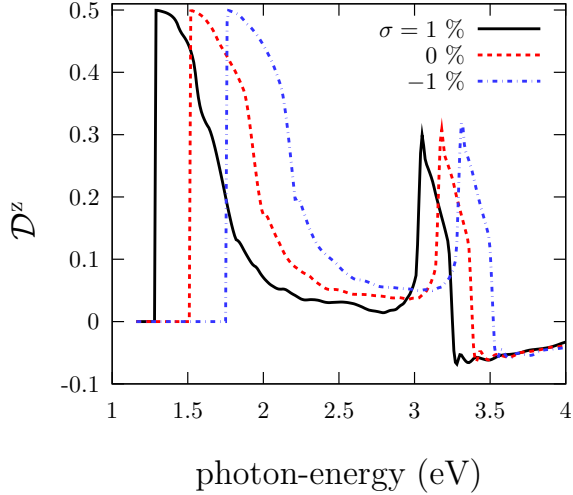}
\caption{(color online) Stress modulation of the DSP, ${\cal D}^{\mathrm{
     z}}$, vs. photon energy for bulk GaAs for three values of strain ($\sigma$).
}
\label{fig_dsp_gaas}
\end{figure}

Finally, in Fig.~\ref{fig_dsp_gaas} we show for GaAs the calculated
${\cal D}^{\mathrm{ z}}$ vs. the photon energy for three values of $\sigma$,
one for expansive stress, one for compressive stress, and the other
one for the unstressed ($\sigma=0$) result.  The unstressed spectrum
shows two positive peaks, one at 1.5~eV, just at the band edge of
GaAs with ${\cal D}^{\mathrm{ z}}=50\%$, and the other at 3.18~eV with
${\cal D}^{\mathrm{ z}}=30\%$. As we expand (compress) the unit cell to
$\sigma=1\%$ ($\sigma=-1\%$) we see that the ${\cal D}^{\mathrm{ z}}$ spectrum
shifts almost rigidly along the energy axis towards lower (higher)
energies, with only a very small change in the intensity of the peak
at 3.1 eV.
 This behavior remains valid for larger values of $|\sigma|$.  For
the unstressed case of GaAs, the $50\%$ value of the ${\cal D}^{\mathrm{ z}}$ has
been confirmed experimentally,\cite{bhatPRB05} and explained
theoretically,\cite{nastosPRB07} thus our calculated results
indicate that either compressive or expansive strain will only move
the onset of the signal. This also shows that the symmetry of the
electronic band structure that leads into the results shown for
$\sigma=0$
remains basically the same as we
apply the stress,\cite{nastosPRB07}  in contrast with Si, where the changes of
${\cal D}^{\mathrm{ z}}$ are readily noticeable.

We have presented a study of optical spin injection rates for stressed
bulk Si and stressed bulk GaAs. 
Both compressive and expansive
stress can effectively modulate the degree of spin polarization in
these materials.  
 For bulk Si, compressive stress shifts the positive peak of
${\cal D}^{\mathrm{z}}$
to higher energies and diminishes the signal  about $20\%$ of its 
value for the unstressed case.
On the other hand, the negative deep remains almost unchanged both in
energy position and magnitude.
Contrary to this behaviour,
for expansive stress we found that the DSP signal is notably
enhanced. 
For $1.5\%$ of volumetric change the line shape of the signal changes
from the one negative deep and one positive peak
of the unstressed case to two positive peaks. One at the band edge
with 50\% of DSP and the other with an almost negligible magnitude.
Thus, expansive strain changes the DSP from 
$-31.5\%$ of the unstressed case to $50\%$.
Further expansion shifts this positive peak
 to lower energies without changing its magnitude.
For bulk
GaAs, compressive and expansive stress rigidly shift the spectrum to
higher or lower energies, respectively, maintaining the band edge peak
signal at $50\%$.  
The results presented in this work show that the application of
stress can be employed to tune the material to a suitable photon
energy and, more importantly, to increase net DSP for the case of Si,
making this material just as efficient as GaAs. We believe this ought
to motivate the experimental verification of the theoretical results
presented here.

We 
acknowledge useful discussions with F. Nastos, J. Sipe and S. Turneaure.
BSM 
acknowledges 
partial support by CONACYT grant 48915-F,
 and CS and JLC scholarships by CONACYT and CONCYTEG.

\bibliography{bibliografia}

\begin{thebibliography}{17}
\bibitem[{not()}]{notamty}
\bibinfo{note}{Essentially, for a 0 to $+8.5\%$ volumetric expansion, the
  signal ${\cal D}^z$ peak shifts from 3.5~eV to 1.5~eV, thus sweeping a large
  portion of the optical range.}
\end{thebibliography}
\end{document}